\documentclass[prd,twocolumn,groupedaddress,showpacs]{revtex4}
\usepackage[dvips]{graphicx}
\usepackage[]{caption}
\usepackage{amsmath}
\usepackage{amssymb}
\def\Dsl{\hbox{/\kern-.6700em\it D}} % D slash
\def\dsl{\hbox{/\kern-.5300em$\partial$}}

\def\eqa{\begin{eqnarray}}
\def\eeqa{\end{eqnarray}}
\def\eq{\begin{equation}}
\def\eeq{\end{equation}}
\def\be{\begin{equation}}
\def\ee{\end{equation}}
\def\bea{\begin{eqnarray}}
\def\eea{\end{eqnarray}}
\def\nn{\nonumber}

\begin{document}
\bibliographystyle{prsty}
\title{On the Decoherence of Primordial Fluctuations During Inflation}

\author{P. Martineau,$^{1)}$}\email[email: ]{martineau@hep.physics.mcgill.ca}
\affiliation{
1) Department of Physics, McGill University, 3600 University Street, Montr\'eal, QC, Canada, H3A 2T8.}

\date{\today}
\begin{abstract}
We study the environment-induced decoherence of cosmological perturbations in an inflationary background. Splitting our spectrum of perturbations into two distinct sets characterized by their wavelengths (super and sub-Hubble), we identify the long wavelength modes with our system and the remainder with an environment. We examine the effects of the interactions between our system and the environment. This interaction causes the long-wavelength modes to decohere for realistic values of the coupling and we conclude that interactions due to backreaction are more than sufficient to decohere the system within 60 e-foldings of inflation. This is shown explicitly by obtaining an analytic solution to a master equation detailing the evolution of the density matrix of the system.
\end{abstract}
\maketitle
\section{Introduction}
As is well known, temperature fluctuations in the CMB and the inhomogeneities that seed structure formation in the universe share a common origin. Both are a result of the scalar metric perturbations produced during inflation. However, these perturbations are of a purely quantum mechanical nature while no cosmological systems of interest (CMB anisotropies, clusters etc.) display any quantal signatures. Presumably, for this to be the case, the primordial density perturbations underwent a quantum-to-classical transition some time between generation during inflation and recombination, when structure first became apparent.

Decoherence is a much studied process (see \cite{Giulini:1996nw} for a comprehensive review). Although not all conceptual issues have been resolved, it is understood that it can occur whenever a quantum system interacts with an ''environment''. In other words, this effect can be said to pervade open systems due to the difficulty of creating a truly closed, macroscopic quantum system. Along with its ubiquity, it is also known to be a practically irreversible process, since the loss of quantum correlations in the system is accompanied by an increase in entropy. 

Early studies of the classicalization of primordial perturbations focussed on intrinsic properties of the system (see, for example \cite{Guth:1985ya},\cite{Polarski:1995jg}). This was made possible by the application of ideas of quantum optics to the theory of cosmological perturbations. Primordial density fluctuations (the scalars as well as the tensors) evolve into a peculiar quantum state - a \textit{squeezed} vacuum state \cite{Grishchuk:1990bj},\cite{Albrecht:1992kf}. By studying the large squeezing limit of these states, it was found that quantum perturbations become indistinguishable from a classical stochastic process. In other words, quantum expectation values in a highly squeezed state are identical to classical averages calculated from a stochastic distribution, up to corrections which vanish in the limit of infinite squeezing.\ The authors of \cite{Polarski:1995jg} refer to this as ''decoherence without decoherence'' while \cite{Kiefer:1998qe} endows the phenomenon with the more technical epithet ''quantum non-demolition measurement''. We emphasize that these works focussed on the classical properties of the states and not on the coherence properties of the system.

As is well understood, in order to study true classicalization, one must consider two distinct aspects of a system. First the quantum states must evolve, in some limit, into a set of states analogous to classical configurations. The second is that these resultant states interfere with each other in a negligible fashion. This last property constitutes decoherence and is equivalent to the vanishing of the off-diagonal elements of the density matrix.

A truly closed gravitational system is a practical impossibility (unless one considers the totality of the universe to constitute the system as in, for example, quantum cosmology). Since the gravitational interaction has infinite range and couples to all sources of energy, interactions with some sort of environment are an inevitability. As such, environmentally induced decoherence must also be present and would play an important role in the classicalization of primordial density fluctuations.

The purpose of the present article is to determine precisely the effects by the ''inflationary environment'' (we will elucidate this notion below) on cosmological perturbations and to study the resultant decoherence. Other authors have also examined this problem (see, for example \cite{Kiefer:1998qe},\cite{Sakagami:1987mp},\cite{Brandenberger:1990bx},\cite{Calzetta:1995ys},\cite{Lombardo:2005iz}) - however, we are the first to present an exact analytic expression for the density matrix with a realistic environment-system interaction.

The paper is organized as follows: in the next section, we review some basic properties of decoherence of which we will make use. After reviewing the quantum theory of cosmological perturbations in section III, we make clear our concept of the environment and motivate some realistic interactions in section IV. Subsequently, we develop necessary formalism which, in section VI, we make use of to demonstrate the classical nature of the system and calculate the decoherence time scale.

\section{Decoherence}

In the present section, we intend to present an extremely (but, hopefully, not exceedingly) terse account of the theory of decoherence. The physics of classicalization is elegant and subtle and a thorough exposition of its finer points would bring us too far afield from the purpose of this article. We confine our attention solely to the cardinal features and disregard any peripheral aspects. The reader unsatisfied by our presentation is encouraged to consult any of a number of excellent reviews of which we mention but a few \cite{Zeh:1999fs},\cite{Zurek},\cite{Giulini:1996nw}.

From an operational perspective, the process of decoherence usually refers to the disappearance of off-diagonal elements of the density matrix. These elements (phase relations) represent the interference of states inherent in any quantum system. Evidently, their disappearance is an integral part of a quantum-to-classical transition.

Having mathematically defined decoherence, we now turn to the physical processes responsible for it. At the heart lies the concept of the open system and the near impossibility of forming a macroscopic closed state. Virtually all realistic systems must interact with an environment of some sort where, by environment, we refer to degrees of freedom which interact with degrees of freedom in our system but which are not witnessed by some observer intent only on the evolution of the system. This leads to the first important characteristic of decoherence - its \textit{ubiquity}.

Next, we come upon the concept of entangled states. Initially, if we disregard all correlations between system and environment, our composite wave function (system $+$ environment) can be expressed as the outer product of the system and environment states (more generally, it will be the outer product of ensembles of states, as is the case when one makes use of density matrices). Though initially factorizable, interactions between the environment-system pair rapidly change this: the total state evolves from the form

\be{|\Psi \rangle\,=\,(\sum_{i} \alpha_{i} |\phi_{i}^{system})\rangle \otimes (\sum_{j} \beta_{j}|\Phi_{j}^{environment}),}\ee
to
\be{|\Psi \rangle\,=\,\sum_{i,j}\gamma_{ij}|\phi_{i}^{system}\rangle |\Phi_{j}^{environment}\rangle},\label{entangled}\ee

\noindent where (\ref{entangled}) represents an entangled state and, as such, is non-factorizable in this basis. Entanglement is key to the whole process for the following reason - an entangled state produces a density matrix which is non-factorizable. The operational equivalent of an observer ignoring the environmental degrees of freedom is to trace out (partial trace) these degrees of freedom. Due to the orthogonality of the environment states, the observer is left with a density matrix which diagonalizes as the states entangle (the fact that the decoherence rate is related to the rate of entanglement has been used to estimate decoherence times. See, for example, \cite{Kiefer:1998qe},\cite{Kiefer:1998jk}). An interesting property of classicalization follows from this - the interference terms are still present, but are unobservable by a ''local'' observer (local in the sense that he only observes the system).

These ''hidden'' interference terms lead us to our next point. By tracing out the environmental degrees of freedom, an observer throws away all the correlation terms, leading to a decrease in the amount of information available in the system - hence, this leads to an increase in the entropy from which we can conclude that decoherence is a practically irreversible process.

The system being decohered, it can only be found in a much smaller subset of the states that were previously allowed - this is what prevents us, in part, from seeing ''Schroedinger's Cat'' states at a macroscopic level. The states that diagonalize the density matrix of the system are referred to as pointer states  \cite{Zurek:1981xq}, and these states remain in the subset of physical states after decoherence. If the evolution of the system is dominated by the self-Hamiltonian of the system, the pointer basis is composed of the eigenstates of the self-Hamiltonian while, if the interaction dominates, the eigenstates of the interaction form the basis \cite{Paz:1998ib}. Pointer states are also those states for which the production of entropy during decoherence is minimized (predictability sieve)\cite{Zurek:1992mv}.

Finally, we conclude with a heuristic view of decoherence. Neglecting certain interacting degrees of freedom in a theory will generally lead to an apparent loss of unitarity. Thus, one should expect a flow of probability out of the system which, in turn, manifests itself as a vanishing of certain elements of the density matrix.

\section{Quantum Perturbations in an Inflationary Universe}

\subsection{The Action for Quantum Perturbations}

We provide in this section an overview of the quantum 
theory of cosmological perturbations in an inflationary background. For a 
more in-depth treatment, the reader is referred to \cite{Mukhanov} or 
\cite{Brandenberger}.

The classical action for an inflationary model is given by (in this and in 
what follows, we set $G=\hbar=1$)
\be
S\,=\,\int{d^{4}x\,\sqrt{-g}(\frac{1}{16\pi}\,R 
+\frac{1}{2}\partial_{\mu}\phi\partial^{\mu}\phi-V(\phi))}.\label{action}
\ee 
\indent If the potential $V(\phi)$ for the matter scalar field
$\phi$ is sufficiently flat and if, in addition, initial 
conditions are chosen for which the kinetic and spatial gradient terms in
the energy density are negligible, this 
action leads to a period of inflation during which the space-time 
background is close to de Sitter
\be{ds^{2}\,=\,(\frac{\alpha}{\eta})^{2}(-d\eta^{2}+(dx^{i})^{2}),}\ee
where $\eta$ is conformal time. 

During the course of inflation, any pre-existing classical fluctuations
are diluted exponentially. However, quantum fluctuations 
are present at all times in the vacuum state of the matter and metric 
fluctuations about the classical background space-time. Their
wavelengths are stretched exponentially, become larger than the
Hubble radius $H^{-1}(t)$ and re-enter the Hubble radius after 
inflation ends. These fluctuations are hypothesized to be the
source of the currently observed density inhomogeneities and 
microwave background anisotropies. In order for this hypothesis to be
correct, the fluctuations must decohere.

The quantum theory of linear fluctuations about a classical
background space-time is a well-established subject (see e.g.
the reviews \cite{Mukhanov} or \cite{Brandenberger}). If the matter
has no anisotropic stress (which is the case if matter is described
by a collection of scalar fields), then a gauge (coordinate system)
can be chosen in which the metric including its (scalar metric) fluctuations
\footnote{We are not considering the vector and tensor metric 
fluctuations. In an expanding background, the vector perturbations
decay, and the tensor fluctuations are less important than the scalar
metric modes.} $(\psi)$ can be written as 
\be
ds^{2} \, = \, (\frac{\alpha}{\eta})^{2}
(-(1+2\psi(x,\eta))\,d\eta^{2} + (1-2\psi(x,\eta))\,(dx^{i})^{2}),
\ee
and the matter including its perturbation $(\delta\phi)$ is
\be{\phi \longrightarrow \phi+\delta\phi(x,\eta).}\ee

The quantum theory of cosmological perturbations is based on the
canonical quantization of the metric and matter fluctuations about
the classical background given by $a(\eta)$ and $\phi(\eta)$. Since
the metric and matter fluctuations are coupled via the Einstein
constraint equations, the scalar metric fluctuations contain only
one independent degree of freedom. To identify this degree of
freedom, we expand the action (\ref{action}) to second order in $\delta\phi$ 
and $\psi$, and combine the terms by making use of the so-called Mukhanov 
variable \cite{Mukhanov2,Lukash}
\be \label{mukhvar}
v \, = \, a(\eta)\,[\delta\phi+\frac{\phi^{\prime}}{\mathcal{H}}\psi],
\ee
in terms of which the perturbed action $S_2$ takes on a canonical form
(the kinetic term is canonical) and the perturbations can hence readily 
be quantized:
\be \label{S_2}
S_2\,=\,\frac{1}{2}\int{d^{4}x\,[v^{\prime 2}-v_{,i}v_{,i}+\frac{z^{\prime\prime}}{z}v^{2}]},
\ee
where $z\,=\,\frac{a\phi^{\prime}}{\mathcal{H}}$, and a prime indicates
a derivative with respect to $\eta$.
This action contains no interaction terms: it represents the evolution of a 
free scalar field with a time-dependent square mass
\be
m^2 \, = -\, \frac{z^{\prime\prime}}{z}, \ 
\ee
propagating in a flat, static spacetime. This action leads directly to a 
well-defined quantum theory via the canonical commutation relations.

The Hamiltonian corresponding to the above action $S_2$ can be written
down in second quantized form:
\be \label{Hamilt}
H \, = \, \int{d^{3}\vec{k}}
[\,k\,(a^{\dag}_{\vec{k}}\,a_{\vec{k}}+a^{\dag}_{-\vec{k}}\,a_{-\vec{k}}+1)
- i\frac{z^{\prime}}{z}\,(a_{\vec{k}}\,a_{-\vec{k}}-h.c.)].
\ee 
The first term in the brackets represents back-to-back harmonic oscillators, 
in phase such that the system has no net momentum.
The second term leads to the ``squeezing'' of the oscillators on scales
larger than the Hubble radius $H^{-1}(t)$ (on these scales
the second term in (\ref{Hamilt})
dominates over the spatial gradient terms coming
from the first term in the equation of motion for $v$). On these scales,
the squeezing leads to an increase in the mode amplitude
\be \label{modeev}
v_k(\eta) \, \sim \ z(\eta) \, \sim \, a(\eta) \, ,
\ee 
where the second proportionality holds if the equation of state
of the background geometry does not change in time. We take this to be the case in our subsequent analysis.

\subsection{Properties of Squeezed States}

There exists an extensive literature on squeezed states. We refer the reader 
to \cite{Schumacher1} and \cite{Schumacher2} for the mathematical properties 
of squeezed states. For their physical interest, we direct the reader to 
\cite{Scully}.

The evolution of a state of a system governed by the Hamiltonian (\ref{Hamilt}) can be described by the following evolution operator:

\be{U\,=\,S(r_{k},\varphi_{k})\,R(\theta_{k}),}\ee

\noindent where

%A two mode squeezed vacuum state is defined by the action of the squeezing 
%operator $S_{\vec{k}}$, an operator derived from the second term in the 
%Hamiltonian which reads
%
\be
S_{\vec{k}}(\eta) \, = \,
\exp[\frac{r_{k}(\eta)}{2}(e^{-2i\varphi_{\vec{k}}(\eta)}a_{-\vec{k}}a_{\vec{k}}-h.c.)], 
\label{squeezing}
\ee

\noindent and 

\be{R(\theta_{k})\,=\,\exp[-i\theta_{k}(a^{\dag}_{k}a_{k}+a^{\dag}_{-k}a_{-k})],}\ee

\noindent where $S(r_{k},\varphi_{k})$ is the two-mode squeeze operator, $R(\theta_{k})$ is the rotation operator, and the real number r is known as the squeeze factor. The rotation operator and the phase $\theta_{k}$ play no important role in what follows hence we ignore them from now on.

The action of the squeezing operator on the vacuum results in squeezed vacuum states
\bea
&S_{\vec{k}}(\eta)|0\rangle\, \equiv |k \rangle \,= \\&\sum^{\infty}_{n=0}
\frac{1}{\cosh(r_{k}(\eta))} 
(-e^{2i \varphi_{k}(\eta)}\,\tanh(r_{k}(\eta)))^{n}\,|n,k;n,-k> \nn.
\eea

The behaviour of the squeezing parameter $r_{k}$ is completely determined by the 
background geometry. The evolution of the squeezing parameters is typically
very complicated, but an exact solution is known in the case of a de Sitter 
background \cite{Albrecht:1992kf}:
\be{r_{k}\,=\,\sinh^{-1}(\frac{1}{2k\eta}),\label{deSittersqueeze}}\ee
\be{\varphi_{k}\,=\,-\frac{\pi}{4}-\frac{1}{2}\arctan(\frac{1}{2k\eta}),}\ee
\be{\theta_{k}\,=\,k\eta+\tan^{-1}(\frac{1}{2k\eta}),}\ee

\noindent where the vacuum state being operated upon corresponds to the Bunch-Davies vacuum.

The squeezing operator has the property of being unitary so that
\be{\langle k|k\rangle\,=\,1.}\ee
Although squeezed states do not provide a basis (as they are overcomplete), they do form an orthogonal 
set of states:
\be{\langle l|k\rangle\,=\,\delta^{l}_{k}.}\ee
This follows from the properties of many particle states.

An important property of squeezed states of which we will make use is the fact that the number of particles in such a state can be expressed entirely in terms of the squeezing parameter via

\be{\langle k|N_{k}|k \rangle\,=\,\sinh^{2}(r_{k}),}\ee

where $N_{k}$is the number operator for the k-mode.
Physically, squeezed states represent states which have minimal uncertainty 
in one variable (high squeezing) of a pair of canonically conjugate variables - the 
uncertainty in the other is fixed by the requirement that the state 
saturates the Heisenberg uncertainty bound. For those states of cosmological interest, we take the squeezing to be in momentum.

For our application, the squeezing parameter will be quite large. As shown in \cite{Grishchuk:1989ss},

\be{r_{k}\,\approx\,\ln(\frac{a(t_{2})}{a(t_{1})}),}\ee

\noindent where $a(t_{1})\,(a(t_{2}))$ is the scale factor at first (second) Hubble crossing. For current cosmological scales, $r_{k}\,\approx 10^{2}$.

\subsection{The Hidden Sector}

An essential ingredient in the theory of decoherence is the presence of unobserved, ''hidden'' degrees of freedom: their interaction with our system degrees of freedom causes the delocalization of the phase relations. In this section, we show that de Sitter space naturally provides us with a hidden sector and that the borderline between the visible and invisible in our theory is naturally given by the Hubble scale.

\begin{figure}
\begin{center}
\includegraphics[width=0.5\textwidth]{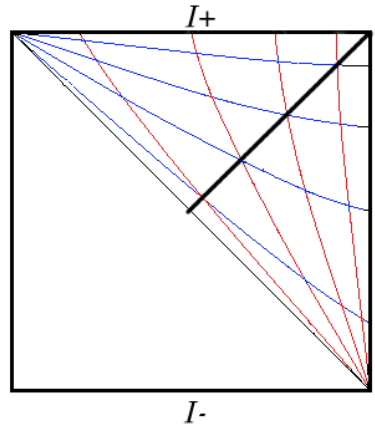}
\caption{The Penrose diagram for de Sitter space in planar coordinates. Note that these coordinates only cover half the spacetime. Blue lines indicate lines of constant t, red lines constant r, and the solid black line represents the horizon.}

\end{center}
\end{figure}

Although de Sitter space is geodesically complete, a geodesic observer will be subject to the effects from both a particle horizon and an event horizon \cite{Hawking},\cite{Spradlin:2001pw}. That the latter constitutes a true event horizon can best be seen by examining the behaviour of null geodesics in Painleve-de Sitter coordinates (see, for example \cite{Parikh:2002qh}), which remain finite across the horizon, in contrast to static coordinates. Specifically, we have

\be{ds^{2}\,=\,-(1-\frac{r^{2}}{l^{2}})\,dt^{2}-2\frac{r}{l}\,dt\,dr+dr^{2}+r^{2}\,d\Omega^{2}.}\ee

Here, l$(=1/H)$ denotes the de Sitter radius. Clearly, setting $r=l$ causes our timelike coordinate to become spacelike (the characteristic feature of an event horizon). Timelike observers that cross from $r-|\epsilon|$ to $r+|\epsilon|$ find themselves incapable of getting back, trapped outside of a sphere of radius l.

Now, if one transforms to the coordinates typically used when discussing inflation (the so-called planar coordinates) and examines the behaviour of timelike geodesics, one finds that all timelike observers originating within the horizon must eventually cross.

The zero-point fluctuations induced by the horizon \cite{Gibbons:1977mu} can be thought of as the seeds for metric perturbations \cite{Hawking:1982cz}, \cite{Bardeen:1983qw}. Heuristically, the horizon can be thought of as a source of thermal radiation with a temperature $H/2\pi$ (in complete analogy with the black hole case). This radiation then produces gravitational metric perturbations, with the same spectrum,  which are stretched out by subsequent cosmological evolution and ultimately lead to the formation of structure in the post-inflationary universe. 

Note, however, that this naive picture is not quite correct - the equation of state of the produced radiation is not thermal \cite{Brandenberger:1983gy}, and including the effects of gravitational back-reaction leads to corrections to the thermal spectrum (this is also true in the black hole case \cite{Parikh:1999mf}). However, our ensuing discussion in no way relies on strict thermality.

We consider our observer to be to the left of the horizon in fig.1. In accord with our discussion above, we take our radiation to be produced at the horizon with a continuous distribution such that a non-vanishing subset of our modes have wavelengths less than l (or $H^{-1}$). It follows that our observer in planar coordinates, due to the event horizon, will be prevented from observing certain radiation modes. We conclude that those modes which are unobservable are those associated with physical wavelengths less than the horizon scale. Of course, gravitational redshifting will cause these modes to stretch and eventually cross the horizon. The point is that particle production is a continuous process and we expect that, at all times, a certain set of modes will be unobservable, and these modes will be associated with physical wavelengths less than $H^{-1}$. As a result of this, decoherence is an inevitability and we define our environment to be a set of modes whose physical momenta are greater than the Hubble scale.

Having identified the modes of the theory which we must trace out, we ask what happens if we trace out additional modes. For example, if an observer was only interested in very low energy modes $(k \ll H)$ he could ignore (or trace out) modes with ($k < H,$ but not $k \ll H$) - surely this would provide an additional source of decoherence as it increases the environment. However, compare this to the case of an observer who is interested in all super-Hubble modes. The second observer would see less decoherence than the first. Decoherence is, after all, an observer dependent effect - an observer who could monitor every degree of freedom in the universe wouldn't expect to see any decoherence. However, our goal is to determine a lower bound on the amount of decoherence as measured by any observer in the ''out'' region of our Penrose diagram. In this case, we trace out only those modes which we must (i.e. all modes on sub-horizon scales) and take our system to be composed of the rest.

\section{Interactions with the Environment}

Key to our investigation of decoherence is the notion of the environment. Such an environment can take on many different guises. As was stated above, we define ours in the following fashion: expanding the background fields (gravity and the inflaton) in terms of fluctuations, we identify our environment with the fluctuations whose wavelengths are less than some cut-off, while our system consists of those wavelengths greater than this cutoff. As explained above, since we are operating in a de Sitter background, the natural scale to pick for the cutoff is the Hubble scale. 

In order to determine the precise form of interactions inherent to a system of cosmological 
perturbations, we expand (\ref{S_2}) to the next order (recall that expanding to second 
order is what led to a free field theory) in the fluctuations, and express the result in 
terms of $v(x,\eta)$. Interactions can either be purely gravitational in nature (backreaction), or they can 
arise in the matter sector through $V(\phi)$, the inflaton potential.

\subsection{Gravitational Backreaction}

To focus on the interactions due to gravitational backreaction, we must expand the 
gravitational action to third order in the amplitude of the perturbations and write down the potential 
in terms of the Mukhanov variable $v$. Expanding to higher order simply introduces more complicated interactions. For our purposes, we restrict our attention to the simplest terms that arise.

In the case where the metric, including
its fluctuation field $\psi$, is given by 
\be
ds^{2} \, = \,a^{2}(\eta)[ -(1+2\psi)d\eta^{2}+(1-2\psi)(dx^{i})^{2}], \ 
\ee
we can expand the Ricci scalar in powers of $\psi$ to obtain

\be{R\,=\, \frac{6\,a(\eta)''}{a^{3}(\eta)(1+2\psi)}\,=\,6\frac{a(\eta)''}{a^{3}(\eta)}(1-2\psi+4\psi^{2}-8\psi^{3}...)}\ee

(where terms with derivatives either temporal or spatial of the $\psi$ have been ignored as they are sub-dominant) from which we can extract our term of interest, $R^{(3)}$,

\be{R^{(3)}\,\equiv\,-48\frac{a(\eta)''}{a^{3}(\eta)}\psi^{3}},\ee
\noindent

\noindent which is the leading order gravitational self-interaction term. Recalling the definition (\ref{mukhvar}) of the Mukhanov variable in a slow-roll inflationary 
background, our potential, expressed in terms of $v$,  becomes (neglecting $\delta \phi$ when substituting
$v$ for $\psi$ and we use the fact that, for our inflationary background, $a(\eta)\,=\,1/(H\eta)$)
\bea \label{above}
V \, &=& \,\frac{1}{16\pi M_{Pl}^{2}} \int{d^{3}x\, \sqrt{-g}R^{(3)}}\\
&=& \frac{1}{M_{Pl}^{2}}\int{}d^{3}x\,a^{4}(\eta)\,\, \frac{4}{\pi} \, \, \frac{a^{\prime\prime}(\eta)}{a^{3}(\eta)}(\frac{\mathcal{H}v}{(\phi)^{\prime}a})^{3} \nonumber \\
&=&\, \frac{3}{\sqrt{2}\pi}\int{}d^{3}x \frac{H^{2}}{M_{Pl}}\frac{a(\eta)}{(2\epsilon)^{3/2}} v^{3},
\eea
so that 

\be{V\, \equiv \, \int{}d^{3}x \lambda v^{3},}\ee

\noindent with

\be{\lambda\,=\,\frac{3}{\sqrt{2}\pi}\frac{H^{2}}{M_{Pl}}\frac{1}{(2\epsilon)^{3/2}} a(\eta)\,=\,a(\eta) \lambda_{0},}\ee

and where we've used the slow roll conditions

\be{H^{2}\,=\,V(\phi)/(3M_{Pl}^{2}), \, \, \, \, 3H \dot{\phi}\,=\,-V',}\ee

\noindent and

\be{\epsilon\,\equiv\, \frac{M^{2}_{Pl}}{2}(\frac{V'}{V})^{2},}\ee

\noindent is one of the slow-roll parameters. Our dimensionful coupling is explicitly time-dependent - this is to be expected since it is associated with a fixed physical scale and our theory (\ref{S_2}) is written entirely in terms of co-moving quantities.

\subsection{Inflaton Interactions}

In addition to the gravitational backreaction terms, there are also interactions due to non-linearities in the matter evolution equation. Consider a model of chaotic inflation with a potential of the form
\be \label{pot}
V = \int{d^{3}x\,\sqrt{-g} \mu \phi^{4}} \, ,
\ee
where $\mu$ is a dimensionless coupling constant.
The perturbations produced during inflation are joint matter and metric
fluctuations. The matter part of the fluctuation, denoted by $\delta\phi$, give 
rise to a cubic term in the interaction potential of the form
\be
V \, \sim \, \int{d^{3}x\, 4 \sqrt{-g} \mu \phi(\delta\phi)^{3}},
\ee
where, in the case of slow-roll inflation, we can treat $\phi$ as a constant. 
Now, writing the potential in terms of the Mukhanov variable (and this time
neglecting $\psi$ in the process of substitution), we have
\be
V \, \sim \, \int{d^{3}x 4 a^{4}(\eta)\mu \phi (\frac{v}{a})^{3}} \,
= \, \int{d^{3}x \, a(\eta)\,4\mu \phi v^{3}},
\ee
so that
\be
\lambda \, = \, 4\mu \phi \,a(\eta) .
\ee \label{coupling}

How do the coupling strengths of the two potentials compare? Taking the ratio of the two, we find
\bea
\frac{\lambda_{inf}}{\lambda_{grav}}\,&=&\, \frac{4\mu\phi a(\eta)}{\frac{3\pi}{\sqrt{2}}\frac{H^{2}}{M_{Pl}} \frac{1}{(2\epsilon)^{3/2}} a(\eta)}\nn \\
\,&=&\, \frac{4\sqrt{2}}{3\pi}(2\epsilon)^{3/2}\frac{\mu\phi}{H^{2}}M_{Pl}\nn \\
\,&=&\, \frac{4\sqrt{2}}{\pi}(2\epsilon)^{3/2}\frac{M_{Pl}^{3}}{\phi^{3}}.\\
\eea

Since the observationally allowed value for $\phi$ at times when fluctuations relevant to current
observations are generated is of the order $10^{-3}M_{pl}$, we find that the gravitational coupling could conceivably dominate depending on the value of $\epsilon$. Since we are only interested in obtaining a lower bound on the decoherence rate, and due to the fact that the exact form of the inflaton potential (along with the initial conditions that determine $\epsilon$) is model dependent, we consider gravitational backreaction to be the main source of decoherence in what follows. Nonetheless, the above demonstrates that inflaton interactions  have the potential to be important.

We couple our system to the environment by writing 

\be{V\,=\,\int{}d^{3}x \lambda v^{3}\, \equiv\, \int{}d^{3}x \lambda v^{2} \varphi,}\ee

\noindent where v now refers only to the expansion of the Mukhanov variable in momenta greater than some cutoff and $\varphi$ is the same field but expanded in terms of the environment modes.

\section{The Density Matrix}

Having determined a candidate interaction between our system and the environment, we now face the task of deriving an appropriate master equation in order to determine the time dependence of our density matrix. Several approaches exist (for example, \cite{master},\cite{FeynmanVernon}) which have been used by a number of authors  - rather, we follow the method of \cite{formalism} which we now review.

We assume that our system of interest is weakly interacting with some environment. The Von Neumann equation for the full density matrix ($\rho$) reads (note that we make use of conformal time. This is due to the fact that our action (\ref{S_2}) is expressed in terms of conformal time)

\be{\frac{d\rho}{d\eta}\,=\,-i[H,\rho],}\label{eq1}\ee

\noindent where H is the total Hamiltonian of the system and can be written as

\be{H\,=\,H_{0}+V},\ee

\noindent where $H_{0}$ is the self-Hamiltonian and V couples the system to the environment. Note that $(\rho)$ denotes the full density matrix for the system \textit{and} the environment.

Switching to the interaction representation (\ref{eq1}) takes on the form

\be{\frac{d\overline{\rho}}{d\eta}\,=\,-i[\overline{V},\overline{\rho}],}\label{eq2}\ee

\noindent where

\be{\overline{\rho}\,=\,\exp(iH_{0}\eta)\,\rho \,\exp(-iH_{0}\eta)},\ee

\noindent with a similar expression for $\overline{V}$.

A perturbative solution of (\ref{eq2}) is found to be given by the following:
\bea
\overline{\rho}&=&\rho_{0}-i\int^{\eta}_{0}{d\tau \,[\overline{V}(\tau),\rho_{0}]}\\ \nn
&(-i)^{2}&\int^{\eta}_{0}{d\tau_{2}}\int^{\tau_{2}}_{0}{d\tau_{1}}[\overline{V}(\tau_{2}),[\overline{V}(\tau_{1}),\rho_{0}]]+...
\eea

Our ultimate goal is to derive an equation of motion for the reduced density matrix ($\rho_{A}=Tr_{B}\rho$, where A denotes the system quantities while B refers to the environment. We use this notation throughout the rest of the paper). To this end, we trace out the environmental degree of freedoms to obtain

\be{\overline{\rho_{A}}\,=\,\rho^{A}_{0}-\int^{\eta}_{0}{d\tau_{2}}\int^{\tau_{2}}_{0}{d\tau_{1}\,Tr_{B}}[\overline{V}(\tau_{2}),[\overline{V}(\tau_{1}),\rho_{0}]]+...}\label{eq3}\ee

Note that the first order term has vanished - this is due to the specific form of our system-environment coupling. Had we used a potential in which an even power of the environment field had appeared, we would have obtained a non-vanishing contribution at this order. Had this been the case, the first order term could have been neglected on the grounds that it would lead to unitary evolution of the system - since our goal is to study the decoherence of the system (a non-unitary process), we can safely ignore such terms.

We find that (see the appendix)

\be{Tr_{B}\, (\overline{V}(x_{1},\eta_{1})\,\overline{V}(x_{2},\eta_{2})\rho)\, =\, \frac{8\pi^{2}\,a^{2}(\eta)}{V\mathcal{H}^{5}}\,\delta(\eta_{1}-\eta_{2})\delta(x_{1}-x_{2}).\label{locality}}\ee

In light of the fact that this equation was derived in the limit of small time intervals, we can approximate the integral in (\ref{eq3}) by the product of the integrand with the time interval, $\eta$. Bringing ($\rho_{0}$) to the left-hand side and dividing both sides by time allows us to write

\be{\frac{\overline{\rho}-\rho_{0}}{\eta}\,=\,\frac{d\overline{\rho}}{d\eta},}\ee

\noindent in the limit of small $\eta$.

As the initial time $(\eta=0)$ is arbitrary, we conclude that our equation for the reduced density matrix may be written as (in terms of physical time)

\be{\frac{d\overline{\rho}(t)}{dt}\,\simeq \,-a(t)\frac{8\pi^{2}\lambda^{2}(t)}{V\mathcal{H}^{5}} \int{d^{3}x}[v^{2},[v^{2},\overline{\rho}(t)]],}\label{master}\ee

\noindent where the details have been relegated to the appendix. V is a normalization volume, $\mathcal{H}\,=\, a(t)H$, with H the physical Hubble scale, and we note that in order to obtain the condition (\ref{locality}), it was necessary to eliminate non-local terms by coarse-graining over scales of order $\mathcal{H}$ in both time and space. 

The differential equation (\ref{master}) is the master equation for our system. In order to proceed, we obtain a matrix representation in the basis of squeezed states. Again, we point out that these do not form a true basis for the Hilbert space (note, however that the use of an overcomplete basis  poses no difficulties when it comes to obtaining representations of the density matrix \cite{Zeh:1999fs}). However, in the limit of large squeezing, squeezed states become orthogonal to other states in the system. Since squeezed states are the "natural" states of the system, we view all other states as being spurious and truncate our Hilbert space so that it contains only the former. Furthermore, as our interactions are small compared to (\ref{Hamilt}), we identify the squeezed states as our pointer basis \cite{Anglin:1995pg}.

Finding a matrix representation of eq.(\ref{master}) is a relatively simple affair - due to the nature of the squeezed states, the expectation value operator $v^{2n}, n \in \mathbb{Z}$ must be diagonal in this (discrete) basis of states. This, along with the identities \cite{Schumacher1}:

\bea
S(r_{k},\varphi_{k})\,a_{\pm k}\,S^{\dag}(r_{k},\varphi_{k})\,&=&\,a_{\pm k}\,\cosh(r_{k})\\ \nn
&+&a^{\dag}_{\mp}\,e^{2i\varphi_{k}}\,\sinh(r_{k}),
\eea
and
\be{S^{-1}(r_{k},\varphi_{k})\,=\,S^{\dag}(r_{k},\varphi_{k})\,=\,S(-r_{k},\varphi_{k}),}\ee

\noindent renders the calculation relatively straightforward. Note that $\langle k|N_{k}|k\rangle\,=\,\sinh^{2}(r_{k})$, where $N_{k}$ is the number operator \cite{Schumacher1}. With this in mind, we find that (\ref{master}) reduces to

\bea
\frac{d\rho_{ij}}{dt}\,&\simeq&\,-a(t) \frac{128\pi^{2}}{V^{2}}\frac{\lambda^{2}(t)}{\mathcal{H}^{5}}\,(\frac{\sinh^{4}(r_{i})}{k_{i}^{2}}+\frac{\sinh^{4}(r_{j})}{k_{j}^{2}}\nn \\
&-&2\frac{\sinh^{2}(r_{i})\sinh^{2}(r_{j})}{k_{i}k_{j}})\rho_{ij} ,\label{master2}
\eea

\noindent where, for simplicity, we've replace the $\cosh^{2}(r)$ terms with $\sinh^{2}(r)$ since we are interested in the limit of large r.

The combination $\frac{\sinh^{2}(r_{i})}{V}\equiv n_{i}(t)=a^{2}(t)n_{i}(0)$ is to be interpreted as the particle density, a quantity which is finite in the thermodynamic limit. Clearly, the decoherence rate increases as the difference between the two momenta increases. For this reason, we take our states of interest to have approximately the same momenta, and the above reduces to 

\bea
\frac{d\rho_{ij}}{dt}\,&\simeq&\,-128\pi^{2} a^{2}(t)\frac{\lambda^{2}_{0} n_{i}^{2}(0)}{H^{5}}\frac{(k_{i}-k_{j})^{2}}{k^{2}_{i}k^{2}_{j}}\rho_{ij},
\eea

\noindent in terms of physical time and co-moving momenta and volume.

A few things are immediately obvious:

1) The diagonal elements suffer no loss of coherence. This actually could have been surmised much earlier from eq.(\ref{eq3}) by noticing that the trace over the system degrees of freedom must vanish.

2) The rate of decoherence grows extremely rapidly. In fact, in order to decohere the system within 60 e-foldings (approximately the minimal time permissible for the duration of inflation), the initial particle density ($n_{0}$) can be as low as $10^{-25}$ particles per Hubble volume. \footnote{In arriving at this estimate, we have considered the case where $k_{i} \approx k_{j}$, $k_{j} \approx H$, $H \approx 10^{-3} M_{Pl}$, $\epsilon \approx 10^{-2}$.}  

3) The particular time $t=0$ for a pair of modes should be taken to correspond to the the time that the shortest of the pair (the higher energy mode) crosses the horizon.

So far, we've argued that a certain sector of the theory is unobservable (thus justifying a minimal amount of tracing), determined an interaction between our visible and invisible sectors, and obtained a lower bound on the parameters of the theory such that decoherence takes place within 60 e-foldings of inflation. The question remains: in a realistic cosmological model, are the parameters of the theory such that decoherence can take place during the inflationary period, and be caused by the leading order gravitational back-reaction term? In other words, is the bound we found satisfied in conventional models?

In order to answer that question, we must obtain the number density of particles in a typical super-Hubble mode at first Hubble crossing.

Consider the square of the substitution we used to obtain our potential in terms of the Mukhanov variable:

\be{v^{2}\,=\,a^{2}(\eta) (\frac{\phi'}{\mathcal{H}})^{2} \psi^{2}.}\label{subs}\ee

To determine the number of particles of the v field in terms of physically meaningful quantities, we must first quantize the Mukhanov field. However, once the theory is quantized, the expression (\ref{subs}) is meaningless - the left-hand side is an operator, while the right is a classical field. In light of this, we follow the usual route \cite{Birrell:1982ix} in semi-classical gravity and replace v with it's vacuum expectation value:

\be{\langle v^{2} \rangle\,=\,a^{2}(\eta) (\frac{\phi'}{\mathcal{H}})^{2} \psi^{2}.}\ee
 
In the limit of large squeezing, we have that

\be{\langle v^{2} \rangle\,=\,\frac{1}{2\pi^{3}}\int{}\frac{d^{3}k}{k}N_{k}(t),}\label{vee}\ee

\noindent where $N_{k}(t)$ is the number of particles in the k-mode at time t, which scales in time as

\be{N_{k}(t) \propto a^{4}(t),}\ee

\noindent where we now consider only physical (as opposed to co-moving as in the previous discussions) quantities. The extra factors of $a(t)$ in the particle number appear because we are now considering the red-shifting of the momenta (see (\ref{deSittersqueeze})). We expect the spectrum to be exponentially suppressed at high (sub-Hubble) momenta: therefore, to a good approximation, the integral in (\ref{vee}) can be taken to be over the infrared sector only. Furthermore, rather than performing the integral over the modes, we reparameterize and integrate over the times which these particular modes first crossed the horizon. In other words, we let

\be{k\,=\,\frac{H}{a(t)},}\ee

\noindent and 

\be{N_{k}(t)\,=\,a^{4}(t) \, N_{H}(0),}\ee

\noindent where, as above, $t=0$ denotes first Hubble crossing for a particular mode. We now have,

\be{\langle v^{2} \rangle \,\simeq \,\frac{2}{\pi^{2}} N_{H}(0)H^{3}\int_{0}^{t_{r}}{}dt a^{2}(t)\,=\,\frac{H^{2}}{\pi^{2}}N_{H}(0)a^{2}(t_{r}),}\ee

\noindent with $t_{r}$ denoting the time of reheating and where we've ignored the time-dependence of the Hubble scale. 

During reheating, the inflaton will undergo periods when it's total energy is dominated by it's kinetic term. So, during reheating, we can make the substitution $\dot{\phi}^{2}\,\simeq\, \rho_{r}$ to obtain

\be{\frac{N_{H}(0)H^{2}}{\pi^{2}}a^{2}(t_{r})\,\simeq\,a^{2}(t_{r})\frac{\rho_{r}\psi^{2}}{H^{2}}.}\ee

We identify $N_{H}H^{3}=n_{H}(0)$ with the number of particles of momentum H per Hubble voume and taking the reheating temperature as H so that $\rho_{r} \simeq H^{4}$. We can now make use of the fact that, observationally, $\psi^{2}\, \approx 10^{-9}$, to deduce that $n_{H}(0)\,\approx 10^{-8}$ particles/Hubble volume. This is well above the lower bound we found. In this case, we find that the modes will decohere approximately 20 e-foldings after crossing the horizon.

\section{Conclusion}

In this paper, we have studied the decoherence of cosmological fluctuations during
a period of cosmological inflation, taking the effects of squeezing into account. We have determined realistic interactions for our system of perturbations and have found that, at the same order, gravitational interactions and  matter (inflaton) interactions are comparable, depending on the scale of inflation and the slow-roll parameter $\epsilon$. Furthermore, we have justified the use of Hubble scale as a cutoff.

Having considered the leading order gravitational correction to the action of quantized cosmological perturbations, we find that super-Hubble modes decohere long before the end of inflation. Of course, we have only obtained a lower bound on the decoherence rate - interactions more complicated than the ones considered here will generally lead to much faster decoherence times \cite{Kiefer:1998jk}.

\section{Appendix: Tracing out the Environment}

In this appendix, we explicitly calculate the partial trace of eq.(\ref{eq3}). 

The expansion of the Mukhanov variable in a spatially flat background takes the form
\be
v \, = \, \frac{1}{(2\pi)^{3/2}}\int{}d^{3}\vec{k}\frac{1}{\sqrt{2|k|}}
(a_{k}e^{-ikx}+a^{\dag}_{k}e^{ikx}),
\ee
and we restrict our attention to modes within a sphere of radius H in momentum space. Since our calculation will be performed in terms of comoving quantities and we take our cutoff to correspond to a fixed physical scale, our cutoff acquires a time dependence of the form

\be{\mathcal{H}\,=\,a(\eta) H.}\ee
Our normalization conventions are as follows:

\be{|k\rangle\,=\,\sqrt{2 E_{k}} a^{\dag}_{k}|0\rangle, \,\,\, \langle k|k^{\prime} \rangle \,=\, (2\pi)^{3} 2E_{k} \delta^{(3)}(k-k^{\prime}),}\nn\ee

\be{[a_{k},a^{\dag}_{k^{\prime}}]\,=\,(2\pi)^{3}\,\delta^{(3)}(k-k^{\prime}).}\ee

\indent The identity operator has the form

\be{1\,=\,\int{}\frac{d^{3}k}{(2\pi)^{3}}\frac{1}{2E_{k}}.}\ee

For simplicity, we ignore the effects of squeezing until the very last. As our initial conditions, we do not take the environment to be in the vacuum - this would be contrary to the basic idea of the generation of inhomogeneities. We take our states to be 2 particle zero-momentum states. Were we to explicitly include the effects of squeezing, we would find that our scattering amplitude $ \langle i|\phi^{n}| j \rangle $ would scales as $a^{n}(\eta)$. Our approach is as follows: we calculate the scattering amplitude for a fixed particle number (2) and, at the last step, include the additional factors of $a(\eta)$ in order to embody the effects of particle production (squeezing). Note that we must take into account squeezing since (\ref{deSittersqueeze}) tells us that all modes in de Sitter space get squeezed.

We take these states to be populated according to a distribution which falls off exponentially in the UV, with temperature parameter $T=\beta^{-1}=\frac{\mathcal{H}}{2\pi}$. In other words

\be{\rho_{env}\,=\,C \exp(-\beta H)},\ee

\noindent where this H refers to the Hamiltonian. The precise form of the distribution is immaterial - after tracing, the only information that the systems retains about the environment is it's ''size'' (the cutoff scale). As another simplification, we take the energy of the state to be dominated by it's momentum. Due to the nature of squeezing and in view of our comments about the distribution, this is a perfectly justifiable assumption. C is a normalization constant which we determine by the condition that the trace of the left hand side of the equation be $\rho_{sys}$ i.e. $Tr_{env} \rho=\rho_{sys}$.
\bea
Tr_{env}\, \rho\,&=&\,\int{}\frac{d^{3}k}{(2\pi)^{3}}\frac{1}{2E_{k}} \frac{\langle k,-k|\rho|k,-k \rangle}{\langle k,-k|k,-k \rangle}\\ \nn
&=&\frac{C}{2\pi^{2}}\rho_{sys}\int^{\infty}_{\mathcal{H}}{}dk\frac{E_{k}}{2}\exp{(-2\beta E_{k})}\\ \nn
&=&\frac{C}{2\pi^{2}}\rho_{sys}(\frac{1}{32\pi^{2}}\mathcal{H}^{2}e^{-4\pi}(1+4\pi))\equiv \rho_{sys}.
\eea

Therefore, we set $C\,\approx\,16\pi^{3}e^{4\pi}/\mathcal{H}^{2}$.

The terms on the right hand side will all have the basic form (aside from the trace of $\rho_{0}$, which is the same as the above):
\bea
RHS\, &=&\,\int{}\frac{d^{3}k}{(2\pi)^{3}}\frac{1}{2E_{k}} \frac{\langle k,-k|v(x)v(x^{\prime})\rho|k,-k \rangle}{\langle k,-k|k,-k \rangle}\\ \nn
%&\propto& \int \frac{d^{3}k}{(2\pi)^{3}}\frac{1}{\delta(0)}[e^{ik(x-x^{\prime})}+e^{-ik(x-x^{\prime})}]e^{-2\beta E_{k}}\\ \nn
&=&  \frac{8\pi}{\delta(0)\mathcal{H}^{4}} \int^{\infty}_{\mathcal{H}} dk\frac{\sin[k (x-x^{\prime})]}{k (x-x^{\prime})}(e^{-i\omega_{k}(\eta-\eta^{\prime})})e^{-2\beta E_{k}},
\eea
\noindent where the delta function arises from the normalization of the states.
Since we are only interested in physics on scales much greater than $\mathcal{H}$, we coarse-grain over time and use the relation

\be{\langle e^{i\omega_{k}(\eta-\eta^{\prime})} \rangle \,\approx \, \frac{\delta(\eta-\eta^{\prime})}{\mathcal{H}}.}\ee

Thus, we find that

\bea
RHS\, = \, \frac{8\pi}{\delta(0)\mathcal{H}^{5}}\delta(\eta-\eta^{\prime})\int^{\infty}_{\mathcal{H}}\,dk \frac{\sin[k(x-x^{\prime})]}{k(x-x^{\prime})}e^{-2\beta E_{k}.}
\eea

Again, as our interest lies in scales such that $\mathcal{H}(x-x\prime) \gg 1$, we perform the substitution

\be{\langle \frac{\sin[k(x-x^{\prime})]}{k(x-x^{\prime})} \rangle \,= \pi \delta(\mathcal{H}(x-x^{\prime})).}\ee

Finally, we obtain

\be{RHS\,\simeq\, \frac{8\pi^{2}}{\delta(0)}\frac{\delta(\eta-\eta^{\prime})\,\delta(x-x^{\prime})}{\mathcal{H}^{5}}a^{2}(\eta).}\ee

Note that we identify $\delta(0)$ with the volume of space and we've included the additional factors of $a(\eta)$ as dicussed above.

\begin{acknowledgments}
This work is supported by funds from McGill University. The author would like to acknowledge useful conversations with Robert Brandenberger and Cliff Burgess. The author is especially indebted to J. Smith for many enlightening discussions.

\end{acknowledgments}

\end{document}